\documentclass[aps,pra,twocolumn,nofootinbib,floatfix,10pt]{revtex4-2}
\usepackage[braket, qm]{qcircuit}
\usepackage{bbold}
\usepackage{appendix}
\usepackage{amsmath}
\usepackage{amssymb}
\usepackage{wasysym}
\usepackage{graphicx}
\usepackage{color,soul}
\usepackage{siunitx}
\usepackage{dsfont}
\usepackage{xcolor}
\usepackage{ulem}
\usepackage[english]{babel}
\usepackage{blindtext}
\usepackage[english,nomargin,inline,marginclue,draft]{fixme}
\pdfpageheight\paperheight
\pdfpagewidth\paperwidth

\usepackage[colorlinks,linkcolor=blue,anchorcolor=blue,citecolor=blue,urlcolor=blue]{hyperref}

\fxusetheme{colorsig}
\FXRegisterAuthor{cg}{acg}{CG}  
\FXRegisterAuthor{th}{ath}{\color{blue}TH}  
\FXRegisterAuthor{ib}{aib}{\color{red}IB} 
\FXRegisterAuthor{sh}{ash}{\color{cyan}SH} 
\FXRegisterAuthor{db}{adb}{\color{green}DB} % now I can ..
\FXRegisterAuthor{ps}{aps}{PS}
\makeatletter
\renewcommand*\FXLayoutInline[3]{%
  {\@fxuseface{inline}\ignorespaces{\color{fx#1}[#3: #2]}}}
\makeatother

\long\def\symbolfootnote[#1]#2{\begingroup%
\def\thefootnote{\fnsymbol{footnote}}\footnotetext[#1]{#2}\endgroup}

\def\nobreakbefore{%
  \relax\ifvmode\else
    \ifhmode
      \ifdim\lastskip > 0pt\relax
        \unskip\nobreakspace
      \else % added to put a ~if no space was typed. (Unclear why it sometimes worked before )
        \nobreakspace
      \fi
    \fi
  \fi
}
\let\oldcite\cite
\renewcommand\cite{\nobreakbefore\oldcite}

\begin{document}
\title{Early warning signals of the tipping point in strongly interacting Rydberg atoms}

\author{Jun Zhang$^{1,2,\textcolor{blue}{\dagger}}$}
\author{Li-Hua Zhang$^{1,2,\textcolor{blue}{\dagger}}$}
\author{Bang Liu$^{1,2}$}
\author{Zheng-Yuan Zhang$^{1,2}$}
\author{Shi-Yao Shao$^{1,2}$}
\author{Qing Li$^{1,2}$}
\author{Han-Chao Chen$^{1,2}$}
\author{Zong-Kai Liu$^{1,2}$}
\author{Yu Ma$^{1,2}$}
\author{Tian-Yu Han$^{1,2}$}
\author{Qi-Feng Wang$^{1,2}$}
\author{C. Stuart Adams$^{3}$}
\author{Bao-Sen Shi$^{1,2}$}
\author{Dong-Sheng Ding$^{1,2,\textcolor{blue}{\star}}$}

\affiliation{$^1$Key Laboratory of Quantum Information, University of Science and Technology of China, Hefei, Anhui 230026, China.}
\affiliation{$^2$Synergetic Innovation Center of Quantum Information and Quantum Physics, University of Science and Technology of China, Hefei, Anhui 230026, China.}
\affiliation{$^3$Department of Physics, Joint Quantum Centre (JQC) Durham-Newcastle, \\Durham University, South Road, Durham DH1 3LE, United Kingdom.}

\date{\today}

\symbolfootnote[2]{J.Z, L.H.Z contribute equally to this work.}
\symbolfootnote[1]{dds@ustc.edu.cn}

\begin{abstract}
    The identification of tipping points is essential for prediction of collapses or other sudden changes in complex systems. Applications include studies of ecology, thermodynamics, climatology, and epidemiology. However, detecting early signs of proximity to a tipping is made challenging by complexity and non-linearity. Strongly interacting Rydberg atom gases offer model systems that offer both complexity and non-linearity, including phase transition and critical slowing down. Here, via an external probe we observe prior warning of the proximity of a phase transition of Rydberg thermal gases. This warning signal is manifested as a deviation from linear growth of the variance with increasing probe intensity. We also observed the dynamics of the critical slowing down behavior versus different time scales, and atomic densities, thus providing insights into the study of a Rydberg atom system's critical behavior. Our experiment suggests that the full critical slowing down dynamics of strongly-interacting Rydberg atoms can be probed systematically, thus providing a benchmark with which to identify critical phenomena in quantum many-body systems.
\end{abstract}
\maketitle

As a result of their inherent interconnectedness and nonlinear dynamics, complex systems have tipping points at which small changes in a specific component or parameter can trigger a disproportionately large response in the entire system \cite{stanley1971phase,goldenfeld2018lectures}. Discovering early warning signals shown in Fig.~\ref{Fig1}(a) before these tipping points are essential steps in enabling complex systems to be managed effectively, which allows us to implement strategies that can avoid or mitigate the detrimental effect of abrupt changes\cite{scheffer2009early,tomen2019functional,barnosky2012approaching,kwong2019identifying,brookes2021critical,xu2023non,dakos2008slowing}. Furthermore, complex systems often show self-organization \cite{ding2019Phase,haken2010information} and emergent properties \cite{wintermantel2021epidemic}, which means that their behavior at the collective level cannot be predicted solely based on an understanding of the individual components. As the interactions and dependencies between the components become stronger near the tipping point, the system will then become more susceptible to abrupt shifts \cite{cael2021abrupt,wade2018terahertz,hagstrom2023phase}. Various factors can contribute to the existence of tipping points in different complex systems; examples include epileptic seizures in medicine \cite{mcsharry2003prediction,anwar2020epileptic}, systemic market crashes in financial networks \cite{may2008ecology,diks2019critical}, climate in nature \cite{van2024physics,lenton2012early,zhang2021predicting}, and phase transitions in physics \cite{carr2013nonequilibrium,eisert2015quantum,onuki2002phase}. Several significant works in the literature have reported on building of theoretical models to predict tipping points \cite{tomen2019functional,wimberger2014nonlinear,hohenberg1977theory}, with examples that include landscape-flux theory \cite{yan2013nonequilibrium,xu2021unifying}, deep learning methods \cite{bury2021deep,maturana2020critical}, and the critical slowing down \cite{lenton2012early,wissel1984universal,zhang2021predicting} which describes how the dynamics of a system change as it approaches a tipping point, by these ways early warning signals are introduced.

The long-range interactions that occur between Rydberg atoms provide us with the ability to simulate large scale systems and study non-equilibrium phase transitions, self-organized criticality, and other phenomena \cite{marcuzzi2014universal,ding2022enhanced,ding2019Phase,schmitt2022quantum,helmrich2020signatures,klocke2021hydrodynamic}. The connection between Rydberg atom systems and complex systems lies in their shared features of the complex interconnectedness, strong nonlinearity and emergent behaviors \cite{gallagher2005Rydberg,de2016intrinsic,urvoy2015strongly,firstenberg2016nonlinear,weimer2008quantum,chialvo2010emergent,grieves2017digital,szabo2017validating}. Therefore, the study of Rydberg atom systems aids in understanding of emergent behaviors and also provides insights into the dynamics of complex systems \cite{yan2023emergent,viteau2012cooperative,browaeys2020many,keesling2019quantum,bernien2017probing,lesanovsky2014out}. However, the complex interconnectedness and nonlinearity, the nonstationary dynamics, and the uncertainty evolution of such systems mean that they are difficult to control and manipulate \cite{bluvstein2021controlling,busche2017contactless,zhang2022photon,sieberer2023universality}. For example, even small changes in the system parameters can lead to significant responses within the entire atomic system, where the behavior of one component can have far-reaching effects on the overall system dynamics, including inducing rich emergent properties and self-organization dynamics \cite{ding2019Phase}. However, no experiments on prediction of the tipping points in Rydberg atom systems have been reported to date, and these systems are worthy of study.

\begin{figure}
\centering
\includegraphics[width=1\linewidth]{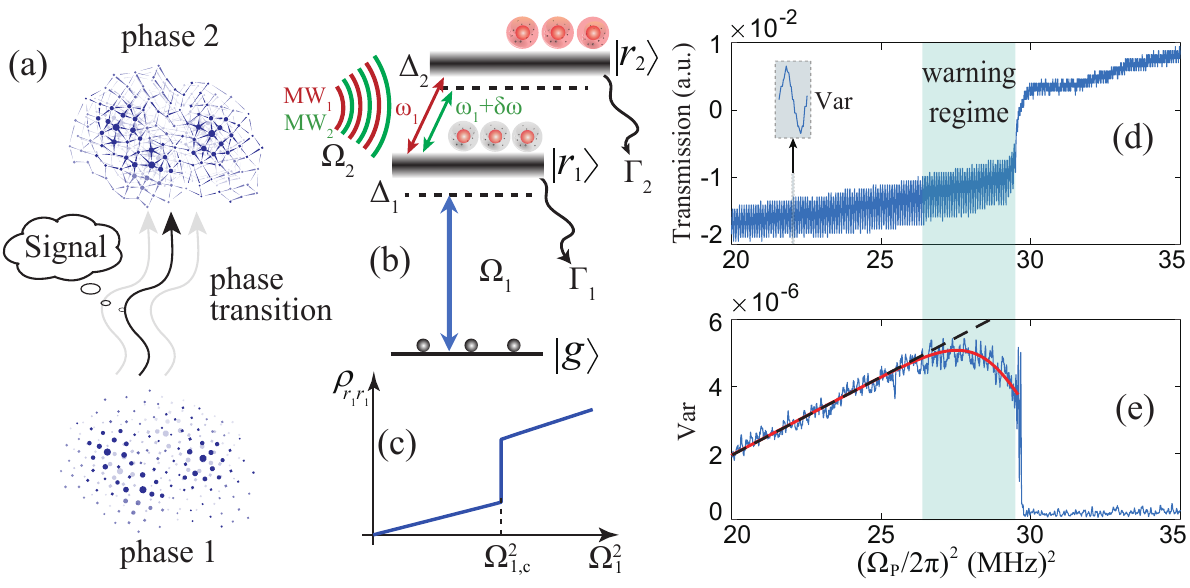}\\
\caption{\textbf{Physical model and early warning signals of the tipping point.} (a) A phase transition in a complex system, in which phase 1 and phase 2 represent the different states of matter below and above the tipping point. (b) Two-level atom model including the ground state $\left| g \right\rangle$ and the Rydberg state $\left| r_1 \right\rangle$, and the atoms are driven by a dual-tone microwave (MW) field. (c) Phase transition of the Rydberg atom system. (d) Measured transmission of the probe beam versus the parameter $\Omega_p^2$ under the MW field driving condition. (e) Measured variance from (d). The early warning signal regime manifest a nonlinear growth property, which is marked as the blue shaded area. The black dotted line is a linear curve. We fitted the trend of the variance using a solid red line, where the fit function is $y = p(x-20)e^{q(x-20)^{\alpha}}$+$s$ ($p$ = -4.44$\times$10$^{-7}$ $\pm$ -4.16$\times$10$^{-10}$, $q$ = -3.89$\times$10$^{-9}$ $\pm$ 3.72$\times$10$^{-10}$, $s$ = 2.01$\times$10$^{-6}$ $\pm$ 1.41$\times$10$^{-9}$, $\alpha$ = 8.50 $\pm$ 0.04).}
\label{Fig1}
\end{figure}

Here, we report an experiment in which the tipping point of interacting Rydberg atoms can be predicted by driving the system using a resonant microwave (MW) electric field. We observe that early warning signals are manifested as a symptom regime in which the variance in the transmission no longer grows linearly. In our system, the signals are extracted from the system's response to MW, whereas the signals in other complex systems are extracted from internal fluctuations\cite{bury2021deep,maturana2020critical,mcsharry2003prediction}. By using a nondestructive measurement method with the electromagnetically induced transparency (EIT), we can coherently map the status of the Rydberg atoms with respect to the probe field. The two metastable states of the non-equilibrium system display different responses because the low-density Rydberg atoms (phase 1) demonstrate relatively small dissipation when compared with that of the high-density Rydberg atoms (phase 2). Interactions make the Rydberg atoms difficult to synchronize under the driving MW field. Therefore, in experiment, we can identify the tipping point at which the system undergoes a phase transition directly. This work provides predictive capabilities with regard to the behavior of complex systems and will enable better early responses to potential tipping points.

%\section*{Results}
%\subsection*{Warning signal model}
%\textit{Warning signal model.---}

\textbf{Warning signal model.}
The inherent interconnectedness and the nonlinear interactions within complex systems make it difficult to predict sudden changes in these systems. Modeling of complex systems can help us to find early warning signals before these sudden changes, as illustrated in Fig.~\ref{Fig1}(a). The Rydberg atom system represents a good platform for study of complex systems because of the strong interactions that occur between the Rydberg atoms, and there is a non-equilibrium phase transition for Rydberg atoms from the phase 1 to the phase 2, as illustrated in Fig.~\ref{Fig1}(b, c).

The Rydberg atom system consists of $N$ interacting three-level 85-Rubidium ($^{85}$Rb) atoms with a ground state $\left| g \right\rangle$, a Rydberg state $\left| r_1 \right\rangle = \left| 50D_{5/2} \right\rangle$ with a decay rate of $\Gamma_1$, and a Rydberg state $\left| r_2 \right\rangle = \left| 51P_{3/2} \right\rangle$ with a decay rate of $\Gamma_2$, as depicted in Fig.~\ref{Fig1}(b). The atoms are driven into the Rydberg state $\left| r_1 \right\rangle$ using a probe beam and a coupling beam. The probe beam excites atoms from the ground state $|g\rangle =|5S_{1/2},F=3\rangle$ to the intermediate excited state $|e\rangle =|5P_{3/2},F=4\rangle$, and the coupling beam drives the transition from $|e\rangle$ to the Rydberg state $|r_1\rangle$. As depicted in Fig.~\ref{Fig1}(b), the intermediate energy level $|e\rangle$ is adiabatically eliminated, thus effective coupling is between the ground state $|g\rangle$ and the Rydberg state $|r_1\rangle$ with a Rabi frequency $\Omega_1$ and a detuning $\Delta_1$. The beams counterpropagate through a thermal Rb vapor, and the EIT spectrum is obtained by detecting the probe beam intensity. Further details about the experiment can be found in the Supplemental Material. 

A dual-tone MW field drives the atoms between the Rydberg states $\left| r_1 \right\rangle$ and $\left| r_2 \right\rangle$, and this MW field takes the form of $E\rm_{MW}$$\cos(\omega_1 t) +E\rm_{MW}\cos{[(\omega_1 + \delta \omega)t]}$, where $E\rm_{MW} = \hbar \Omega_2 /\mu\rm_{MW}$ represents the MW field amplitude [$\Omega_2$ is Rabi frequency of the MW field and $\mu\rm_{MW}$ is the transition dipole moment], $\omega_1$ is the resonant frequency with respect to the transition between the two Rydberg states $\left| r_1 \right\rangle$ and $\left| r_2 \right\rangle$, and $\delta \omega$ is the MW frequency detuning from resonance. After application of the rotation wave approximation, this dual-tone MW field turns into a periodic driving field with frequency $\delta \omega$. 

When the MW field is off and the intensity profile of the probe laser is scanned, a non-equilibrium phase transition occurs that is characterized by a sudden increase in the population of the Rydberg atoms $\rho_{r1r1}$, as shown in Fig.~\ref{Fig1}(c). When the dual-tone MW field is on, an oscillation effect occurs in the transmission spectrum. The MW field serves as a driving in this case and the spectrum represents the response of the Rydberg atoms to this driving, as illustrated in Fig.~\ref{Fig1}(d) .

To acquire the warning signals before the tipping point, we calculate the variance using the moving window algorithm. The formula is
\begin{equation}
Var=\frac{1}{m-1} \sum_{i=1}^m\left|a_i-\mu\right|^2
\end{equation}
where $a_i$ is the transmission intensity of probe beam, $m$ is the data length of a window, and $\mu$ is the mean value in the window.

When the Rabi frequency of the probe field $\Omega_p$ is low and the system is outside the warning regime, which is shaded in blue in Fig.~\ref{Fig1}(d), the Rydberg atom system is a linear system and its response to the MW field is also linear, as indicated by the linear growth observed in the variance as the square of the Rabi frequency $\Omega_p^2$ increases. However, near the sharp jump in the transmission, there is a warning regime region colored blue. The variance in the region deviates from the previously linear increase, represents a warning signal.

To model the above process, we consider the effect of interactions-induced dissipation, in which interactions between distinct Rydberg atoms could cause spatially dependent phase shifts and increases the total relaxation rate \cite{de2016intrinsic, ding2019Phase}. The detuning and the decay terms are replaced by $\Delta_{1} \rightarrow \Delta_1 - V \rho_{r_1r_1}^\beta$ and $\Gamma_{r_1} \rightarrow \Gamma_{r_1} - V \rho_{r_1r_1}^\epsilon$ \cite{carr2013nonequilibrium,ding2019Phase,ding2022enhanced,de2016intrinsic,lee2012collective} in the mean-field approximation, where $\beta$ and $\epsilon$ are exponent factors. By solving the master equation and expanding the solution $\rho_{r_1r_1}$ as a function of $\Omega_1$ to the second order, the transmission against $(\Omega_1/2\pi)^2$ is obtained. Here, $\beta=1$ and $\epsilon=3$ are selected to model the results better.

To observe the full dynamics near the tipping point, we fit the trend of variance using the function $y = p(x-20)e^{q(x-20)^{\alpha}}$+$s$ [see the solid red line in Fig.~\ref{Fig1}(e)], where $p$, $q$ and $s$ are the coefficients, and $\alpha$ represents the critical exponent which describes the features of the critical slowing down in the non-equilibrium system. The critical exponent $\alpha$ characterizes the system's behavior near the tipping point and describes how it scales as system approaches the tipping point.

\begin{figure}[t]
\centering
\includegraphics[width=1\linewidth]{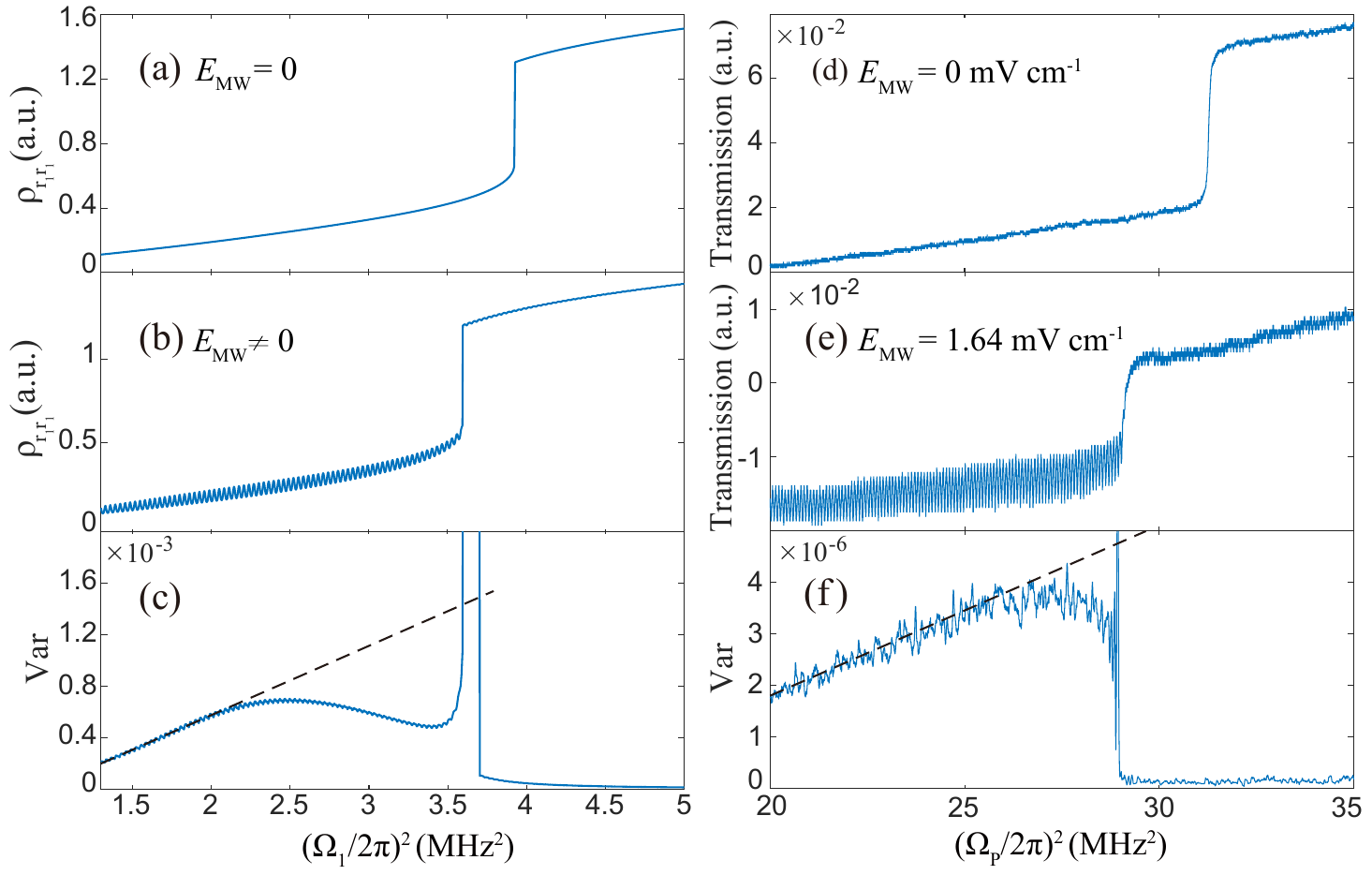}\\
\caption{\textbf{Theoretical and experimental results without and with the driving MW field.} (a), (b) Theoretical simulations of the population $\rho_{r_1r_1}$ of the Rydberg atoms with different electric field amplitudes: (a) $E_{\mathrm{MW}}$ = 0 $\mathrm{mV/cm}$ and (b) $E_{\mathrm{MW}}$ $\neq$ 0 $\mathrm{mV/cm}$. (c) Variance calculated from (b). (d), (e) Experimental transmission spectra under the different MW field amplitudes: (d) $E\rm_{MW}$ = 0 $\mathrm{mV/cm}$ and (e) $E\rm_{MW}$ = 1.64 $\mathrm{mV/cm}$. (f) Measured variance from (e). }
\label{Fig2}
\end{figure}

%\subsection*{Transmission spectra under the dual-tone MW field driving conditions}
\textbf{Transmission spectra under the dual-tone MW field driving conditions.} We apply the dual-tone MW field to the theoretical model and simulate the transmission, where the MW field takes the form of $\Omega_2=\mu\rm_{MW} E_{\mathrm{MW}}(1+\mathrm{Cos}(\delta{\omega}t))/ \hbar$. The simulated results for $\rho_{r_1r_1}$ versus $(\Omega_1/2\pi)^2$ are shown in Fig.~\ref{Fig2}(a-b), where panel (a) corresponds to the case of the phase transition without the MW field ($E_{\mathrm{MW}}=0$) and panel (b) shows the case with the dual-tone MW field ($E_{\mathrm{MW}}\neq 0$). Fig.~\ref{Fig2}(c) shows the variance that was calculated from Fig.~\ref{Fig2}(b), with nonlinear behavior in the region where $2.2 \text{ } (\mathrm{MHz})^2<(\Omega_1/2\pi)^2<3.5 \text{ } (\mathrm{MHz})^2$. This nonlinear behavior serves as the early warning signals, but the signals are difficult to extract without the MW field, shown in Figs.~\ref{Fig2}(a) and (d). Applying the dual-tone MW field amplifies the signals and makes the interactions-induced dissipation easier to detect.

By scanning the probe field intensity, we can observe the non-equilibrium phase transition in the Rydberg atom system, as illustrated in Fig.~\ref{Fig2}(d). To subtract the early warning signals before the tipping point is reached, we experimentally apply a dual-tone MW electric field to drive the transition between the two different Rydberg states $|50D_{5/2}\rangle$ and $|51P_{3/2}\rangle$, with a center frequency of $2\pi \times 17.041$ GHz and a frequency difference of $\delta \omega $=$2\pi \times 0.2$ MHz. The MW field induces a small ac Stark shift that moves the tipping point and varies the Rydberg atom population. The MW field causes the tipping point to move forward when compared with the case without the MW field, and the oscillation appears in the transmission spectrum in Fig.~\ref{Fig2}(e). By calculating the variance from the oscillated transmission, we observe a nonlinear variance when $(\Omega_p/2\pi)^2>26 \text{ } (\mathrm{MHz})^2$, indicating the existence of an early warning signal, as shown in Fig.~\ref{Fig2}(f). The experimental results show good agreement with the expectations of the theoretical simulations.

From the results shown in Fig.~\ref{Fig2}(f), the measured variance reduces nonlinearly when $(\Omega_p/2\pi)^2>$ 27.5\text{ } $(\mathrm{MHz})^2$. The increased $\Omega_p$ adds a number of excited Rydberg atoms, which makes the system dimension higher and then induces a strong critical slowing down effect. In a relatively low-dimensional system where the number of degrees of freedom is limited, localized interactions between neighboring Rydberg atoms are dominant. As a result, the critical slowing down process in a low-dimensional system is primarily influenced by these localized interactions. In this case, all the excited Rydberg atoms in the system can be synchronized rapidly, thus allowing faster recovery from the perturbation. In contrast, a high-dimensional system has large numbers of interacting Rydberg atoms, which results in a rich interconnectedness within the system. In this system, the critical slowing down effects are amplified because of the extended reach and the influence of the long-range interactions. Perturbations take longer times to propagate through the system, and the process of recovery from disturbances thus becomes slower.

\begin{figure}[b]
\centering
\includegraphics[width=1\linewidth]{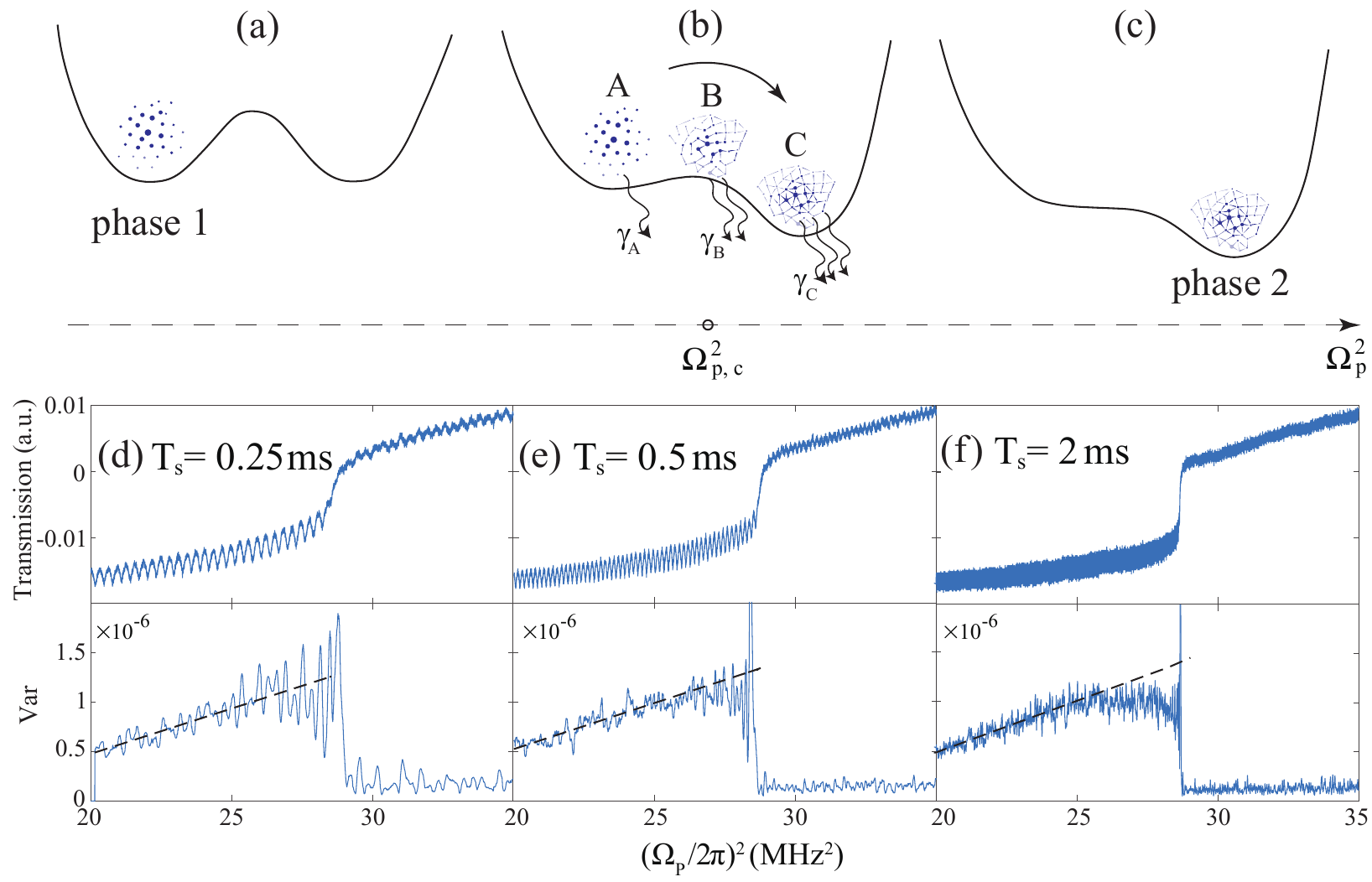}\\
\caption{\textbf{Time evolution of the system during the phase transition process.} (a)-(c) Double well model used to describe the phase transition and the critical slowing down phenomenon. With increasing ${\Omega_{\mathrm{p}}}^2$, the system is initially in phase 1 (${\Omega_{\mathrm{p}}}^2 < {\Omega_{\mathrm{p,c}}}^2$, where ${\Omega_{\mathrm{p,c}}}^2$ is the tipping point), and it then evolves from state A to state B when close to the tipping point, before finally making the transition to phase 2 ((${\Omega_{\mathrm{p}}}^2 > {\Omega_{\mathrm{p,c}}}^2$), state C). In this process, the system states A, B, and C have different interaction-induced dissipations $\gamma_A$, $\gamma_B$, and $\gamma_C$, respectively, where $\gamma_C > \gamma_B > \gamma_A$. (d)-(f) Measured transmission spectra and their corresponding variances within scanning times of (d) 0.25 ms, (e) 0.5 ms, and (f) 2 ms.}
\label{Fig3}
\end{figure}

%\subsection*{Transmission spectra versus scanning time}
\textbf{Transmission spectra versus scanning time.}
The non-equilibrium phase transition in the Rydberg atom system can be modeled using a double-well transition \cite{marcuzzi2014universal}, with each well standing for one bi-stable state, as depicted in Fig.~\ref{Fig3}(a)-(c). The dissipation of phase 2 is greater than that of phase 1 because of the interactions among the Rydberg atoms. 

When the population of Rydberg atoms is increased by $\Omega_{p}^2$, a phase transition occurs from the left well to the right well that crosses the barrier of the double well, as illustrated in Fig.~\ref{Fig3}(b). Near the tipping point $\Omega_{p,c}^2$, the critical slowing down phenomenon occurs; this term refers to an effect where the system's response time to the driving MW field becomes slower (here, the MW field is regarded as a perturbation to the system). After the phase transition, the system is then in the phase 2, and the strong interactions between the Rydberg atoms make the system’s dissipation so large that the variance drops to a very low level. 

%Here, the dissipation is regarded as a perturbation to the system and the MW field is regarded as a driving to the system.  which exhibits a non-linear growth in variance with the intensity of probe beam

To observe the slower dynamics that occur near the tipping point, we changed the scanning time (T$_{s}$) of the triangular wave loaded onto the acousto-optic modulator (AOM) and recorded the dynamics within the warning regime. As shown in Fig.~\ref{Fig3}(d)-(f), we found that the warning regime gradually becomes apparent as the scanning rate is gradually reduced.

A fast scan accumulates a small number of Rydberg atoms because of the small interval time. This causes the system to have low dissipation until the phase transition occurs, which means that the warning regime is almost invisible [see Fig.~\ref{Fig3}(d) and Fig.~\ref{Fig3}(e) for the scanning times of T$_{s}$= 0.25 ms and T$_{s}$= 0.5 ms, respectively]. As the light intensity increases, the system’s response to the MW field follows the increasing light intensity well.

As the scanning time increases, the system can accumulate increasing numbers of Rydberg atoms for a relatively large time interval, and the dissipation of these interacting Rydberg atoms competes with the coherent driving of the laser and the MW field. The dissipation becomes dominant when the system approaches the tipping point, as indicated by the transition process $A\rightarrow B \rightarrow C$ shown in Fig.~\ref{Fig3}(b). This causes the variance to diverge from the linear increasing trend, and thus the warning regime gradually becomes visible [see Fig.~\ref{Fig3}(f) for the scanning time of T$_{s}$ = 2 ms]. 

\begin{figure}[t]
\centering
\includegraphics[width=1\linewidth]{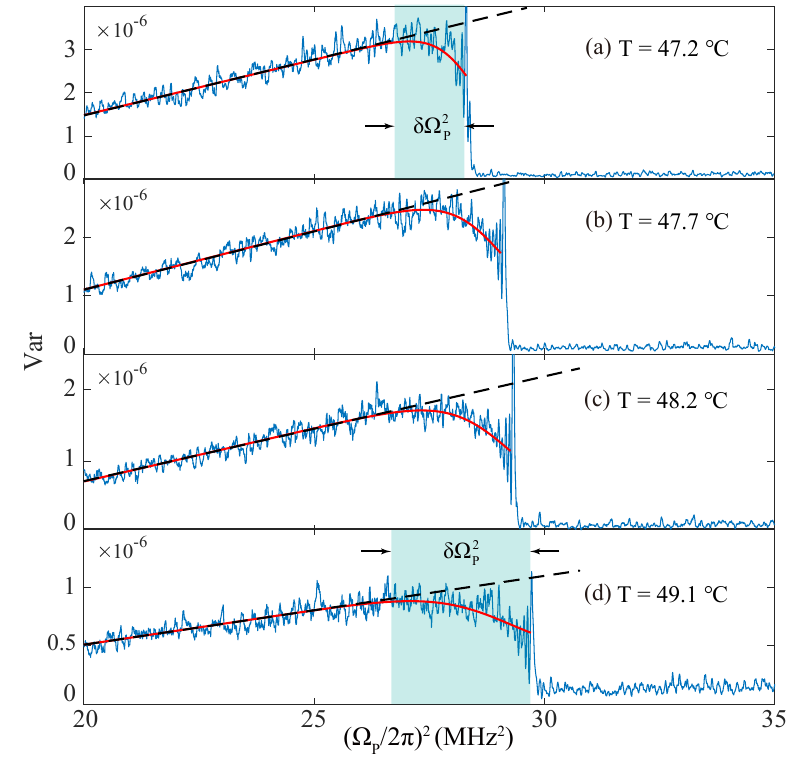}\\
\caption{\textbf{Variance measured at different temperatures.} We measured the different variance characteristics at temperatures of (a) T=47.2$^{\circ}$C, (b) T=47.7$^{\circ}$C, (c) T=48.2$^{\circ}$C, and (d) T=49.1$^{\circ}$C. The early warning signal regimes shown in (a) and (d) are marked as the blue shaded areas for comparison. The solid red lines are the fittings with the function $y = p(x-20)e^{q(x-20)^{\alpha}}$+$s$, where $\alpha$ = 16.46 $\pm$ 0.22 for T=47.2$^{\circ}$C, $\alpha$ = 12.69$\pm$ 0.10 for T=47.7$^{\circ}$C, $\alpha$ = 10.89 $\pm$ 0.07 for T=48.2$^{\circ}$C, and $\alpha$ = 8.59 $\pm$ 0.07 for T=49.2$^{\circ}$C.}
\label{Fig4}
\end{figure}

%\subsection*{Transmission spectra at different temperatures}
\textbf{Transmission spectra at different temperatures.}
In our experiments, we also studied how the interaction between the Rydberg atoms affects the warning signals. A higher temperature leads to a higher interaction level, and vice versa. We changed the atomic density by varying the temperature and measured the variance of the oscillated transmission. Figure~\ref{Fig4}(a)-(d) show the measured results. We found that the variance before the phase transition was decreasing gradually as the temperature increased.
For comparison, we define the width of the warning regime as $\delta \Omega_\mathrm{p}^2$. We found that the early warning signal regime shown in Fig.~\ref{Fig4}(d) was broader than that shown in Fig.~\ref{Fig4}(a), as given by ($\delta \Omega_\mathrm{p}/2\pi)^2 = 1.5 \text{ } (\mathrm{MHz})^2$ at T = 47.2$^{\circ}$C and ($\delta \Omega_\mathrm{p}/2\pi)^2 = 3.0 \text{ } (\mathrm{MHz})^2$ at T = 49.1$^{\circ}$C. Therefore, it is easier to identify the early warning signal at higher temperatures.This is explained by the fact that the interaction-induced dissipation causes the system to be difficult to synchronize when driven by the MW field, and it then leaves the linear response regime. This enhanced critical slowing down process in high-dimensional systems can be attributed to the increased system complexity and the interdependences among the Rydberg atoms.

The early warning signals of the Rydberg atom systems under different spatial structure conditions and scales [from their distinct densities] can also be probed, and the measured exponent $\alpha$ varies with the temperature [e.g., for T = 47.2$^{\circ}$C, $\alpha$ = 16.46 $\pm$ 0.22 in Fig.~\ref{Fig4}(a), but $\alpha$ = 8.59 $\pm$ 0.07 for T = 49.2$^{\circ}$C in Fig.~\ref{Fig4}(d)], thus indicating distinct critical dynamics before occurrence of a phase transition. This confirms that the early warning signal arises from the interactions between the Rydberg atoms, even when the system is still in the phase 1. This finding is consistent with the fact that the non-equilibrium phase transition in the Rydberg atom system occurs because of the interactions among a few local atoms, which eventually spreads to involve the atoms on a global scale. Using the MW field, we can detect small numbers of locally interacting Rydberg atoms, even without causing a phase transition to occur in the system.

%\section*{Discussion}
\textbf{Discussion.}
It should be noted that the measured variance in our experiment was reduced in the early warning regime, which differs from the increased variance that was reported previously \cite{bury2021deep,maturana2020critical,mcsharry2003prediction}. In those cases, the variance reveals the fluctuations of internal variables of the system; while in our system, the variance represents the strength of oscillation under external MW field driving. In this way, the interaction-induced dissipation before the tipping point is related to the amplitude of oscillation, thus allowing us to characterize the full dynamics of the critical slowing down process. The early warning signals for the phase transition cause us to conclude that the interaction-induced dissipation appears first, and the phase transition occurs after that. This is also consistent with the results from our mean-field method, in which the exponent factor $\epsilon$ was greater than $\beta$.

The critical slowing down process of the Rydberg atoms is a type of nonlinear dynamic behavior that can provide technologies and methods to enable study of other nonlinear systems. The interactions within the Rydberg atom system can help us to understand and study the relationships and the coupling effects among the different factors when studying the early warning signals that occur in other complex systems, including atmospheric systems, ecological systems, and socio-economic systems. Most importantly, the technological method for amplification of the signal of the critical slowing down process of Rydberg atoms can provide a new perspective to enable the development and application of prediction strategies and techniques for use in other complex systems. In addition, the collective behavior and self-organization phenomena among the atoms when driven by the MW field can help us to study the essence and the mechanisms of the self-organization phenomena during intervention by external factors, in fields as diverse as traffic flow, urban planning, and immune systems.

In summary, we have observed the full dynamics of the early warning signals at the tipping point for strongly interacting Rydberg atoms. Because of the distinct dissipation of the metastable states of the Rydberg atom system, the responses before and after the tipping point can be probed successfully using the external MW field. The critical slowing down process causes the response to the MW field to deviate from linear growth before the tipping point, which shows good agreement with the theoretical observations. This experiment has helped us to study the critical slowing down that occurs near the criticality of the Rydberg atom system, and provides a deeper understanding of the critical slowing down behavior in other complex systems. This may also promote practical developments in detection of early warnings before catastrophic events in other complex systems.

\hspace*{\fill}

We acknowledge funding from the National Key R\&D Program of China (Grant No. 2022YFA1404002), the National Natural Science Foundation of China (Grant Nos. U20A20218, 61525504, and 61435011), the Anhui Initiative in Quantum Information Technologies (Grant No. AHY020200), the major science and technology projects in Anhui Province (Grant No. 202203a13010001). 

J.Z implemented the physical experiments with L.H.Z. D-S.D. and J.Z conducted the theoretical model. All authors contributed to the experiment preparation and data discussions regarding the results. D-S.D. conceived the idea and support this project.

\normalem
\bibliography{ref}

\maketitle

\onecolumngrid

\appendix
\newpage

\begin{center}
\textbf{Supplemental Material for \lq \lq Early warning signals of the tipping point in strongly interacting Rydberg atoms"}

\end{center}

% 手动插入作者信息
\begin{center}
{Jun Zhang$^{1,2,\textcolor{blue}{\dagger}}$, Li-Hua Zhang$^{1,2,\textcolor{blue}{\dagger}}$, Bang Liu$^{1,2}$, Zheng-Yuan Zhang$^{1,2}$, Shi-Yao Shao$^{1,2}$, \\Qing Li$^{1,2}$, Han-Chao Chen$^{1,2}$, Zong-Kai Liu$^{1,2}$, Yu Ma$^{1,2}$, Tian-Yu Han$^{1,2}$, \\Qi-Feng Wang$^{1,2}$, C. Stuart Adams$^{3}$, Bao-Sen Shi$^{1,2}$, Dong-Sheng Ding$^{1,2,\textcolor{blue}{\star}}$}
\end{center}
% 手动插入单位信息
\begin{center}
\textit{$^1$Key Laboratory of Quantum Information, University of Science and Technology of China, \\Hefei, Anhui 230026, China.} \\
\textit{$^2$Synergetic Innovation Center of Quantum Information and Quantum Physics, \\University of Science and Technology of China, Hefei, Anhui 230026, China.} \\
\textit{$^3$Department of Physics, Joint Quantum Centre (JQC) Durham-Newcastle, \\Durham University, South Road, Durham DH1 3LE, United Kingdom.}
\end{center}
\symbolfootnote[2]{J.Z, L.H.Z contribute equally to this work.}
\symbolfootnote[1]{dds@ustc.edu.cn}

\begin{center}
\textbf{Theoretical model}
\end{center}

A toy numerical model with an interaction-dependent energy shift and energy broadening can reveal the physical mechanism of our experiment. 

The master equation is
\begin{equation}
    \partial_t \hat{\rho} = i [\hat{H},\hat{\rho}] + \mathcal{L}_{r_1}[\hat{\rho}] + \mathcal{L}_{r_2}[\hat{\rho}]
\end{equation}
where $\hat{H}$ is the Hamiltonian and $\mathcal{L}_{\{r_1,r_2\}}$ is the Lindblad jump operator.
The Hamiltonian of the system after adiabatic elimination of the intermediate energy level between  $\left| g \right\rangle$ and $\left| r_1 \right\rangle$ is given by

\begin{align*}
    \hat{H} & =\frac{1}{2}\sum_{i}\left(\Omega_{1}\sigma_{i}^{gr_1}+\Omega_{2}\sigma_{i}^{r_1r_2}+h.c.\right)+\sum_{i}\left(\Delta_{1}n_{i}^{r_1}+\Delta_{2}n_{i}^{r_2}\right)\\ & +\frac{1}{2}\sum_{i\neq j}(V_{ij}^{r_1r_2}n_{i}^{r_1}n_{j}^{r_2}+V_{ij}^{r_2r_1}n_{i}^{r_2}n_{j}^{r_1}+V_{ij}^{r_1r_1}n_{i}^{r_1}n_{j}^{r_1}\\ & +V_{ij}^{r_2r_2}n_{i}^{r_2}n_{j}^{r_2})
\end{align*}
where $\sigma_{i}^{gr_1}$ and $\sigma_{i}^{r_1r_2}$ represent the $i$-th atom transition between $\left| g \right\rangle$ and $\left|  r_1 \right\rangle$ and $\left| r_1 \right\rangle$ and $\left|  r_2 \right\rangle$; $n_{i}^{r_1,r_2}$ are the population operators of the two Rydberg energy levels $\left|  r_1 \right\rangle$ and $\left|  r_2 \right\rangle$; $V_{ij}^{r_1r_2}$ and $V_{ij}^{r_2r_1}$ represent the interactions between the Rydberg atoms in states $\left|r_{1,2}\right\rangle$ and $\left|r_{2,1}\right\rangle$, respectively; and $V_{ij}^{r_1r_1}$ and $V_{ij}^{r_2r_2}$ represent the interactions between the Rydberg atoms in states $\left|  r_1 \right\rangle$ and $\left|  r_2 \right\rangle$, respectively.  

The Lindblad jump terms are:
\begin{equation}
    \mathcal{L}_{r_1} = (\Gamma_{1}/2) \sum_i (2 \hat{\sigma}_i^{r_1 g} \hat{\rho} \hat{\sigma}^{ gr_1}_i - \{\hat{n}_i^{r_1},\hat{\rho}\}),
\end{equation}
\begin{equation}
    \mathcal{L}_{r_2} = (\Gamma_{2}/2) \sum_i (2 \hat{\sigma}_i^{r_2 r_1} \hat{\rho} \hat{\sigma}^{ r_1r_2}_i - \{\hat{n}_i^{r_2},\hat{\rho}\}),
\end{equation}
where these terms represent the decay process from the Rydberg state $\left| r_1 \right\rangle$ to the ground state $\left| g \right\rangle$, and the Rydberg state $\left| r_2 \right\rangle$ to the Rydberg state $\left| r_1 \right\rangle$.  

Note here that the mean-field approximation is used and that the interaction terms ($V_{ij}^{r_1r_2}$, $V_{ij}^{r_2r_1}$, $V_{ij}^{r_1r_1}$, and $V_{ij}^{r_2r_2}$) are replaced by a single atom immersed in a field formed by these interactions between other atoms and this atom. Therefore, the task of solving the dynamic many-body problem is reduced to solving the dynamic problem for one single atom in that field. Besides, before the non-equilibrium phase transition occurs, the interaction between Rydberg atoms dissipates the atoms, which makes the dissipation ahead of the phase transition (the deviation from linear growth of the variance before the tipping point as shown in Fig.1(e).). To model this process, we introduce a density-dependent dissipation in the mean-field approximation, and the detuning and the decay are replaced such that $\Delta_{1} \rightarrow \Delta_1 - V \rho_{r_1r_1}^\beta$ and $\Gamma_{r_1} \rightarrow \Gamma_{r_1} - V \rho_{r_1r_1}^\epsilon$ \cite{carr2013nonequilibrium,ding2019Phase,ding2022enhanced,de2016intrinsic,lee2012collective}, respectively, where $\beta$ and $\epsilon$ are exponent factors. By solving the master equation using the mean-field approximation and expanding the solution $\rho_{r_1r_1}$ as a function of $\Omega_1$ to the second order. Here, $\Omega_1 \propto \Omega_p \Omega_c / \Delta$, where $\Omega_p$ and $\Omega_c$ are Rabi frequencies of probe beam and coupling beam, respectively, and the $\Delta$ represents the detuning of intermediate state. This allows us to obtain the transmission against the intensity of the probe field [proportional to $\Omega_p^2$], which exhibits the same trend as $(\Omega_1/2\pi)^2$. Here, $\beta=1$ and $\epsilon=3$ to enable better modeling of the results.

\begin{figure*}[t]
    
    \centering
    \includegraphics[width=1\linewidth]{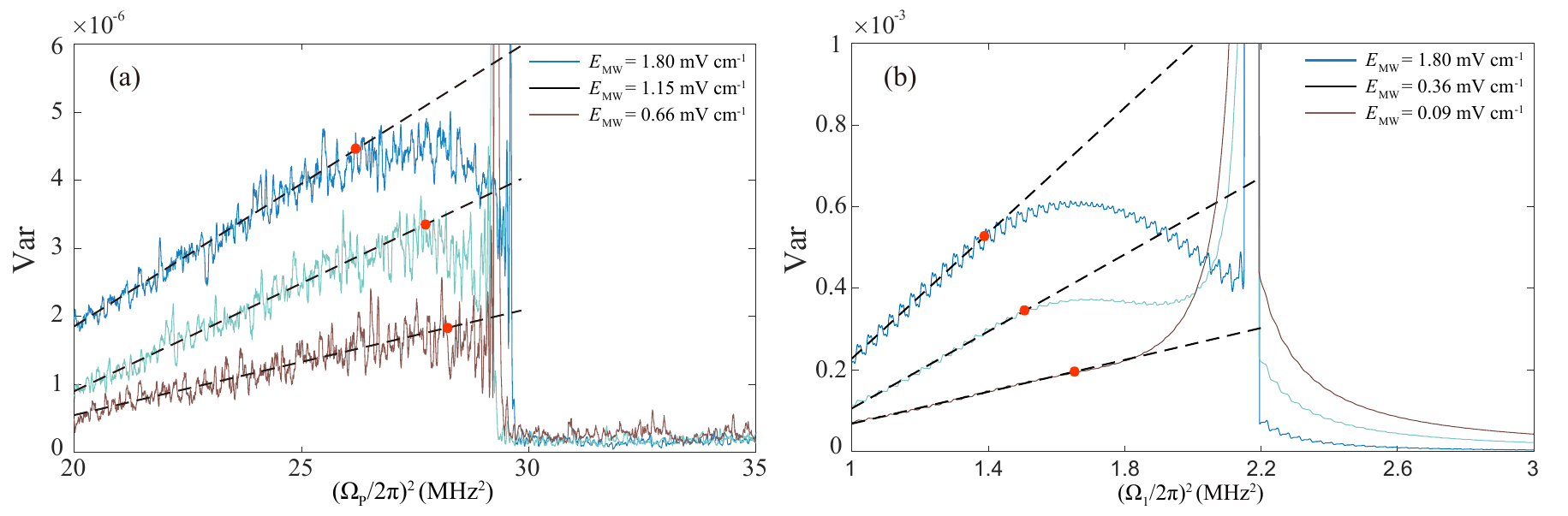}
    \caption{\textbf{Trend of variance at different MW field intensities.} (a) Variance measured at the different MW field intensities of $E_{\mathrm{MW}}$=1.80 mV/cm (blue), $E_{\mathrm{MW}}$=1.15 mV/cm (light green), and $E_{\mathrm{MW}}$=0.66 mV/cm (brown). To provide a better comparison of the signals at the different MW intensities, we normalized the variance of the transmission for easier reading. The variance at 0.66 mV/cm is amplified by a factor of four and the variance at 1.15 mV/cm is amplified by a factor of 1.5, but the variance at 1.80 mV/cm is not amplified. (b) Theoretical simulations at the different modulation strengths of $E_{\mathrm{MW}}$ = 1.8 mV/cm (blue), $E_{\mathrm{MW}}$ = 0.36 mV/cm (light green), and $E_{\mathrm{MW}}$ = 0.09 mV/cm (brown), where we normalized the simulated results using an amplification factor of 10 at $E_{\mathrm{MW}}$ = 0.36 mV/cm and an amplification factor of 20 at $E_{\mathrm{MW}}$ = 0.09 mV/cm; the variance at $E_{\mathrm{MW}}$ = 1.8 mV/cm was not amplified. The red points correspond to the points where the variance diverged from the linear function in each case. In these processes, the tipping points for the different MW field intensities are calibrated to be in the same position for ease of comparison by slightly adjusting the coupling detuning.}
    \label{Fig5}
\end{figure*}
%\subsection*{Transmission spectra versus the MW field intensity}

\begin{center}
\textbf{Transmission spectra versus the MW field intensity}
\end{center}

To study the early warning signals under various MW field intensities, we alter the MW field intensity and record the corresponding variance; see the results in Fig.~\ref{Fig5}(a). When $E\rm_{MW}$ = 0.66 mV/cm, the variance is almost linear versus the probe field intensity. During this process, the MW field intensity reflects the magnitude of the perturbation. For a small perturbation, the Rydberg atoms can follow the driving MW field even though the system is very close to the tipping point, which means that the warning regime is unclear. When we increase the MW field amplitude, the variance diverges from the linear curve, as shown in the cases where $E\rm_{MW}$ = 1.15 mV/cm and $E\rm_{MW}$ = 1.80 mV/cm. The system then cannot follow the driving action of the MW field, which makes the warning regime become more apparent. 

These results can be explained as the process by which the system is synchronized or not when driven by the MW field. The non-interacting Rydberg atoms can cooperate well and are further synchronized with the MW field, while the interacting Rydberg atoms in the system are difficult to synchronize because of the effect of the critical slowing down process. We also simulated the results with different MW field intensities, with results as shown in Fig.~\ref{Fig5}(b). The simulated results show the same trend as the experimental measurements.

\begin{center}
\textbf{Experimental setup}
\end{center}

We used a two-photon transition scheme to excite the alkali metal atoms (rubidium 85) from the ground state to the Rydberg state. The probe beam ($\omega_{p} \approx$ 950 µm) is red-detuned by $2\pi \times $ 60 MHz, driving the atoms from the ground state $|g\rangle =|5S_{1/2},F=3\rangle$ to the intermediate excited state $|e\rangle =|5P_{3/2},F=4\rangle$, and the coupling beam ($\omega_{c} \approx$ 350 µm, $\Omega_{\mathrm{c}}/2\pi$ = 4.72 MHz) then drives the transition from $|e\rangle$ to the Rydberg state $|r_1\rangle =|50D_{5/2}\rangle$. We use a MW electric field to drive the RF transition between two different Rydberg states $|50D_{5/2}\rangle$ and $|51P_{3/2}\rangle$, and the MW electric field used in our experiment was generated by a RF source and a horn antenna. A 780 nm laser beam was first divided into two beams using a beam splitter, and the intensity of one of these beams was detected using a power meter. This method allows us to monitor the intensity of the experimental beam in real time. The other beam was split by a beam displacer (BD) into a probe beam and an identical reference beam, and both beams propagate in parallel through a heated Rb cell at a temperature of 47$^{\circ}$C. The probe beam is then overlapped with a counterpropagating coupling beam, thus constituting the EIT process. In the experiments, we caused the AOM to work in a linear range. We loaded a triangular wave signal that was generated using a signal generator (RIGOL DG4102) onto the AOM. In this way, we generated a probe beam with linearly increasing intensity ($(\Omega_{p}/2\pi)^2$ from 19.83 MHz$^2$ to 35.22 MHz$^2$). We used the Pound-Drever-Hall method to lock the probe field frequency, and the frequency of coupling beam is also fixed. By slightly adjusting the detuning of the coupling beam, we observed a sudden jump in probe transmission, which corresponds to the non-equilibrium transition from the phase 1 to the phase 2 for the Rydberg atom systems. When the dual-tone MW field was switched on, an oscillation effect occurs in the transmission spectrum. The transmission signals are detected using a differential photodetector and the data are collected through an oscilloscope. 

\begin{center}
\textbf{Generation and calibration of the MW fields}
\end{center}

The MW electric field used in our experiment was generated using an RF source (Ceyear 1465F-V) and a horn antenna. The RF source operates within the frequency range from DC to $2\pi \times 40$ GHz, and the horn antenna was located near the Rb cell. To calibrate the electric field amplitude $E_{\mathrm{MW}}$, we used the RF source to load an MW electric field signal with a frequency of $2\pi \times 17.041$ GHz, which is resonant with the transition of two Rydberg states, $|50D_{5/2}\rangle$ and $|51P_{3/2}\rangle$. Experimentally, we scanned the coupling detuning to obtain the EIT spectrum, and Autler-Townes splitting appeared in the EIT spectrum when the RF source was turned on. The MW electric field amplitude $E_{\mathrm{MW}}$ was calibrated based on the frequency interval $\Delta w$ of the Autler-Townes splitting.
\begin{equation}
    E_\mathrm{MW}=\frac{h \Delta w}{\mu_d} \cdot \frac{\lambda_p}{\lambda_c}
\end{equation}

where $h$ is Planck’s constant, $\mu_d$ is the transition dipole moment of the two Rydberg states, and $\lambda_p$ and $\lambda_c$ are the wavelengths of the probe field and the coupling field, respectively. To obtain the strength of the dual-tone MW field, we measured the signal from the RF source when connected directly to a spectrum analyzer (Ceyear 4024f) for calibration.

\end{document}